\documentclass[nofootinbib,showpacs,preprintnumbers,amsmath,amssymb,floatfix]
{revtex4}
\usepackage{epsf}

\begin{document}


\title{Spin structure function $g_1$ at small $x$ and arbitrary $Q^2$:
Total resummaion of leading logarithms vs Standard Approach}

\vspace*{0.3 cm}

\author{B.I.~Ermolaev}
\affiliation{Ioffe Physico-Technical Institute, 194021
  St.Petersburg, Russia}
\author{M.~Greco}
\affiliation{Department of Physics and INFN, University Rome III,
Rome, Italy}
\author{S.I.~Troyan}
\affiliation{St.Petersburg Institute of Nuclear Physics, 188300
Gatchina, Russia}

\begin{abstract}
The Standard Approach (SA) for description of the structure
function $g_1$ combines the DGLAP evolution equations and Standard
Fits for the initial parton densities. The DGLAP equations
describe the region of large $Q^2$ and large $x$, so there  are
not theoretical grounds to exploit them at small $x$. In practice,
extrapolation of DGLAP into the region of large $Q^2$ and small
$x$ is done with complementing DGLAP with special, singular ($\sim
x^{-a}$) phenomenological fits for the initial parton densities.
The factors $x^{-a}$ are wrongly believed to be of the
non-perturbative origin. Actually, they mimic the resummation of
logs of $x$ and should be expelled from the fits when the
resummation is accounted for. Contrary to SA, the resummaton of
logarithms of $x$ is a straightforward and natural way to describe
$g_1$ in the small-$x$ region. This approach can be used at both
large and small $Q^2$ where DGLAP cannot be used by definition.
Confronting this approach and SA demonstrates that the singular
initial parton densities and  the power $Q^2$-corrections (or at
least a sizable part of them) are rather not real physical
phenomena but the artefacts caused by extrapolating DGLAP into the
small-$x$ region.
\end{abstract}


\maketitle

\section{Introduction}

The Standard Approach (SA) for description of the structure
function $g_1$ involves the DGLAP evolution equations\cite{dglap}
and Standard
 Fits\cite{fits} for the initial parton densities $\delta q$ and $\delta g$.
 The fits are defined from phenomenological considerations at $x\sim
 1$ and $Q^2 = \mu^2 \sim 1$GeV$^2$.
The DGLAP equations are one-dimensional, they describe the $Q^2$
-evolution only, converting $\delta q$ and $\delta g$ into the
evolved distributions $\Delta q$ and $\Delta g$. They represent
$g_1$ at the region \textbf{A}:
\begin{equation}\label{rega}
\textbf{A:}~~~~~ Q^2 \gg \mu^2,~~~~~x \lesssim 1.
\end{equation}

The $x$ -evolution is supposed to come from convoluting $\Delta q$
and $\Delta g$ with the coefficient functions $C_{DGLAP}$.
However, in  the leading order $C_{DGLAP}^{LO}=1$; the NLO
corrections account for one- or two- loop contributions and
neglect higher loops. It is the correct approximation in the
region \textbf{A} but becomes false in the region \textbf{B:}
\begin{equation}\label{regb}
\textbf{B:} ~~~~~Q^2 \gg \mu^2,~~~~~x \ll 1
\end{equation}

where contributions $\sim \ln^k(1/x)$ are large and should be
accounted for to all orders in $\alpha_s$. $C_{DGLAP}$  do no
include the total resummation of leading logarithms of $x$ (LL),
so there  are not theoretical grounds to exploit DGLAP at small
$x$. However regardless of that, SA extrapolates DGLAP into the
region \textbf{B}, invoking special fits for $\delta q$ and
$\delta g$. A general structure of such fits (see
Refs.~\cite{fits}) is as follows:
\begin{equation}\label{fits}
\delta q = N x^{-a} \varphi(x)
\end{equation}
where $N$ is a normalization constant; $a > 0$, so $x^{-a}$ is
singular when $x \to 0$ and $\varphi (x)$ is regular in $x$ at $x
\to 0$. In Ref.~\cite{egtfit} we showed that the role of the
factor $x^{-a}$ in Eq.~(\ref{fits}) is to mimic accounting for the
total resummation of LL performed in Refs~\cite{egtns, egts}.
Similarly to LL, the factor $x^{-a}$ provides the steep rise to
$g_1$ at small $x$ and sets the Regge asymptotics for $g_1$ at $x
\to 0$, with the exponent $a$ being the intercept. The presence of
this factor is very important for extrapolating DGLAP into the
region \textbf{B:} When the factor $x^{-a}$ is dropped from
Eq.~(\ref{fits}), DGLAP stops to work at $x \lesssim 0.05$ (see
Ref.~\cite{egtfit} for detail). Accounting for the LL resummation
is beyond the DGLAP framework because LL come the phase space
violating the base of DGLAP: the DGLAP -ordering
\begin{equation}\label{order}
\mu^2  < k^2_{1~\perp} < k^2_{2~\perp}<...< Q^2~
\end{equation}
for  the ladder partons. LL can be accounted only when the
ordering Eq.~(\ref{order}) is lifted and all $k_{i~\perp}$ obey
\begin{equation}\label{llorder}
\mu^2 < k^2_{i~\perp} < (p+q)^2 \approx (1-x)2pq \approx 2pq~
\end{equation}
at small $x$. Replacing Eq.~(\ref{order}) by Eq.~(\ref{llorder})
leads inevitably to the change of the DGLAP parametrization
\begin{equation}\label{dglapparam}
\alpha_s^{DGLAP} = \alpha_s(Q^2)
\end{equation}
by the alternative parametrization of $\alpha_s$ given by
Eq.~(\ref{a}). This parametrization was obtained in
Ref.~\cite{egta} and was used in Refs.~\cite{egtns,egts} in order
to find explicit expressions accounting for the LL resummation for
$g_1$ in the region \textbf{B}. Obviously, those expressions
invoke the fits for the initial parton densities without the
singular factors $x^{-a}$. Let us note that replacement of
Eq.~(\ref{order}) by Eq.~(\ref{llorder}) brings a more involved
$\mu$ -dependence to $g_1$. Indeed, Eq.~(\ref{order}) makes
contributions of gluon ladder rungs be infrared (IR) stable, with
$\mu$ acting as a IR cut-off for the lowest rung and $k_{i~\perp}$
playing the role of the IR cut-off for the
 $i+1$-rung. In contrast, Eq.~(\ref{llorder}) implies that $\mu$ acts
as the IR cut-off for every rung.

 Besides the
regions \textbf{A} and \textbf{B}, it i necessary to know $g_1$ in
the region \textbf{C}:
\begin{equation}\label{regc}
\textbf{C:} ~~~~~Q^2 < \mu^2,~~~~~x \ll 1~
\end{equation}
because this region is studied experimentally by the COMPASS
collaboration. Obviously, DGLAP cannot  be exploited here.
Alternatively, in Refs.~\cite{egtsmq,egthtw} we obtained
expressions for $g_1$ in the region \textbf{C}. In
Ref.~\cite{egtsmq} we showed that $g_1$ practically does not
depend on $x$ at small $x$, even at $x \ll 1$. Instead, it depends
on the total invariant energy $2pq$. Experimental investigation of
this dependence is extremely interesting because according to our
results $g_1$, being positive at small $2pq$, can turn negative at
greater values of this variable. The position of the turning point
is sensitive to the ratio between the initial quark and gluon
densities, so its experimental detection  would enable to estimate
this ratio.  In Ref.~\cite{egthtw} we analyzed the power
contributions $\sim 1/(Q^2)^k$ to $g_1$ usually attributed to
higher twists. We proved that a great amount of those corrections
have a simple perturbative origin and resummed them. Therefore,
the genuine impact of higher twists can can be estimated only
after accounting for the perturbative $Q^2$ -corrections.

\section{Description of $g_1$ in the region \textbf{B}}

The total resummation of the double-logarithms (DL) and single-
logarithms of $x$ in the region \textbf{B } was done in
Refs.~\cite{egtns,egts}. In particular, the non-singlet component,
$g_1^{NS}$ of $g_1$ is
\begin{equation}
\label{gnsint} g_1^{NS}(x, Q^2) = (e^2_q/2) \int_{-\imath
\infty}^{\imath \infty} \frac{d \omega}{2\pi\imath }(1/x)^{\omega}
C_{NS}(\omega) \delta q(\omega) \exp\big( H_{NS}(\omega)
\ln(Q^2/\mu^2)\big)~,
\end{equation}
with new coefficient functions  $C_{NS}$,
\begin{equation}
\label{cns} C_{NS}(\omega) =\frac{\omega}{\omega -
H_{NS}^{(\pm)}(\omega)}
\end{equation}
and anomalous dimensions $H_{NS}$,
\begin{equation}
\label{hns} H_{NS} = (1/2) \Big[\omega - \sqrt{\omega^2 -
B(\omega)} \Big]
\end{equation}
where
\begin{equation}
\label{b} B(\omega) = (4\pi C_F (1 +  \omega/2) A(\omega) +
D(\omega))/ (2 \pi^2)~.
\end{equation}
 $ D(\omega)$ and $A(\omega)$ in Eq.~(\ref{b}) are
expressed in terms of  $\rho = \ln(1/x)$, $\eta =
\ln(\mu^2/\Lambda^2_{QCD})$, $b = (33 - 2n_f)/12\pi$ and the color
factors
 $C_F = 4/3$, $N = 3$:

\begin{equation}
\label{d} D(\omega) = \frac{2C_F}{b^2 N} \int_0^{\infty} d \rho
e^{-\omega \rho} \ln \big( \frac{\rho + \eta}{\eta}\big) \Big[
\frac{\rho + \eta}{(\rho + \eta)^2 + \pi^2} \mp
\frac{1}{\eta}\Big] ~,
\end{equation}

\begin{equation}
\label{a} A(\omega) = \frac{1}{b} \Big[\frac{\eta}{\eta^2 + \pi^2}
- \int_0^{\infty} \frac{d \rho e^{-\omega \rho}}{(\rho + \eta)^2 +
\pi^2} \Big].
\end{equation}
$H_{S}$  and $C_{NS}$ account for DL and SL contributions to all
orders in $\alpha_s$. Eq.~(\ref{a}) and (\ref{d}) depend on the IR
cut-off $\mu$  through variable $\eta$. It is shown in
Refs.~\cite{egtns,egts} that there exists an Optimal scale for
fixing $\mu$: $\mu \approx 1$ Gev for $g_1^{NS}$  and $\mu
\approx 5$
 GeV for $g_1^s$.  The arguments in favor of existence of
 the Optimal scale were
 given in Ref.~\cite{egthtw}.  Eq.~(\ref{gnsint})
 predicts that  $g_1$ exhibits the power behavior in  $x$ and $Q^2$ when $x \to 0$:
\begin{equation}
\label{gnsas}g_1^{NS} \sim \big(Q^2/x^2\big)^{\Delta_{NS}/2},~
g_1^{S} \sim \big(Q^2/x^2\big)^{\Delta_{S}/2}
\end{equation}
where the non-singlet and singlet intercepts are $\Delta_{NS} =
0.42,~\Delta_{S} = 0.86$ respectively. However the asymptotic
expressions (\ref{gnsas}) should be used  with great care:
According to Ref.~\cite{egtfit}, Eq.~(\ref{gnsas}) should not be
used at $x \gtrsim 10^{-6}$. So, Eq.~(\ref{gnsint}) should be used
instead of Eq.~(\ref{gnsas}) at available small $x$. Expressions
accounting the total resummation of LL for the singlet $g_1$ in
the region \textbf{B} were obtained in Ref.~\cite{egts}. They are
more complicated than Eq.~(\ref{gnsint}) because involve two
coefficient functions and four anomalous dimensions.

\section{Description of $g_1$ in the region \textbf{C}}

Region \textbf{C} is defined in Eq.~(\ref{regc}). It includes
small $Q^2$, so there are not large contributions
$\ln^k(Q^2/\mu^2)$ in this region. In other words, the DGLAP
ordering of Eq.~(\ref{order}) does not make sense in the region
\textbf{C }, which makes impossible exploiting DGLAP here. In
contrast, Eq.~(\ref{order}) is not sensitive to the value of $Q^2$
and therefore the total resummation of LL does make sense in the
region \textbf{C}. In Ref.~\cite{egtsmq} we suggested that the
shift
\begin{equation}\label{shift}
Q^2  \to  Q^2 + \mu^2
\end{equation}
would allow for extrapolating our previous results (obtained in
Refs.~\cite{egtns,egts} for $g_1$ in the region \textbf{B}) into
the region \textbf{C}. Then in Ref.~\cite{egthtw} we proved this
suggestion. Therefore, applying Eq.~(\ref{shift}) to $g_1^{NS}$
leads to the following expression  for $g_1^{NS}$ valid in the
regions\textbf{ B} and\textbf{ C}:
\begin{equation}
\label{gnsbc} g_1^{NS}(x+z, Q^2) = (e^2_q/2) \int_{-\imath
\infty}^{\imath \infty} \frac{d \omega}{2\pi\imath }
\Big(\frac{1}{x+z}\Big)^{\omega} C_{NS}(\omega) \delta q(\omega)
\exp\big( H_{NS}(\omega) \ln\big((Q^2+\mu^2)/\mu^2\big)\big)~,
\end{equation}
where $z = \mu^2/2pq$. Obviously, Eq.~(\ref{gnsbc}) reproduces
Eq.~(\ref{gnsint}) in the region \textbf{B}. Expression for
$g_1^S$ looks similarly  but  more complicated, see
Refs.~\cite{egtsmq,egthtw} for detail. Let us notice that the idea
of considering DIS in the small-$Q^2$ region through the shift
Eq.~(\ref{shift}) is not new. It was introduced by Nachtmann in
Ref.~\cite{nacht}  and used after that by many authors (see e.g.
\cite{bad}), being based on different phenomenological
considerations. On the contrary, our approach is based on the
analysis of the Feynman graphs contributing to $g_1$.

\section{Prediction for the COMPASS experiments}

\begin{figure}
\begin{center}
\begin{picture}(240,140)
\put(0,0){
\epsfbox{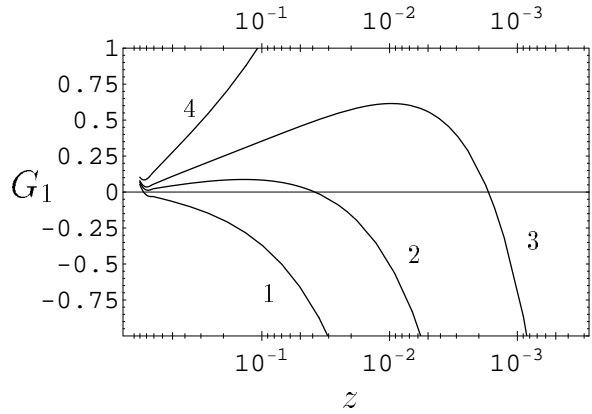}}
\end{picture}
\end{center}
\caption{$G_1$ evolution with decreasing $z=\mu^2/2(pq)$ for
different values of ratio $r= \delta g/\delta q$: curve 1 - for
$r=0$, curve 2 - for $r= -5$~, curve~3 -for $r = -8$ and curve~4
-for $r = -15$.} \label{fig1}
\end{figure}

The COMPASS collaboration now measures the singlet $g_1^S$ at $x
\sim 10^{-3}$ and $Q^2 \lesssim 1$~GeV${^2}$, i.e. in the
kinematic region beyond the reach of DGLAP. However, our formulae
for $g_1^{NS}$ and $g_1^S$ obtained in  Refs.~\cite{egtsmq,egthtw}
cover this region. Although expressions for singlet and
non-singlet $g_1$ are different, with formulae for the singlet
being much more complicated, we can explain the essence of our
approach, using Eq.~(\ref{gnsbc}) as an illustration. According to
results of \cite{egts}, $\mu \approx 5$ GeV for $g_1^S$, so in the
COMPASS experiment $Q^2 \ll \mu^2$. It means, $\ln^k(Q^2 + \mu^2)$
can be expanded into series in $Q^2/\mu^2$, with the first term
independent of $Q^2$:
\begin{equation}\label{gssmq}
g_1^S(x+z, Q^2,\mu^2) = g_1^S(z,\mu^2) + \sum_{k=1} (Q^2/\mu^2)^k
E_k(z)
\end{equation}
where $E_k(z)$ account for the total resummation of LL of $z$ and
\begin{equation}\label{cs}
g_1^S(z,\mu^2) = (<e^2_q/2>) \int_{-\imath \infty}^{\imath \infty}
\frac{d \omega}{2\pi \imath} \big(1/z \big)^{\omega}
\big[C_{S}^q(\omega)\delta q(\omega) + C_{S}^g(\omega) \delta
g(\omega)  \big],
\end{equation}
so that $\delta q(\omega)$ and $\delta g(\omega)$ are the initial
quark and gluon densities respectively and $C_{S}^{q,g}$ are the
singlet coefficient functions. Explicit expressions for
$C_{S}^{q,g}$ are given in Refs.~\cite{egts,egtsmq}. The standard
fits for $\delta q$ and $\delta g$ contain singular factors $\sim
x^{-a}$ which mimic the total resummation of leading logarithms of
$x$. Such a resummation leads to the expressions for the
coefficient functions different from the DGLAP ones. After that
the singular factors in the fits can be dropped and the initial
parton densities can be approximated by constants:
\begin{equation}\label{deltaqg}
\delta q \approx N_q ~~~~\delta g \approx N_g~,
\end{equation}
so, one can write
\begin{equation}\label{g1num}
g_1(Q^2 \ll \mu^2) \approx (<e^2_q>/2) N_q G_1(z)
\end{equation}
with
\begin{equation}
\label{g1q0} G_1 =  \int_{-\imath \infty}^{\imath \infty} \frac{d
\omega}{2\pi\imath }(1/z)^{\omega} \big[C_S^q + (N_g/N_q C_S^g)
\big]. ~
\end{equation}
Obviously, $G_1$ depends on the ratio $N_g/N_q$. The results for
different values of the ratio $r=N_g/N_q$, $G_1$ are plotted in
Fig.~\ref{fig1}. When the  gluon density is neglected, i.e. $N_g =
0$ (curve~1), $G_1$ being positive at $x \sim 1$, is getting
negative very soon, at $z < 0.5$ and falls fast with decreasing
$z$. When $N_g/N_q = -5$ (curve~2), $G_1$ remains positive and not
large until $z \sim 10^{-1}$, turns negative at $z \sim 0.03$ and
falls afterwards rapidly with decreasing $z$ . This turning point
where $G_1$ changes  its sign is very sensitive to the magnitude
of the ratio $r$~. For instance, at $N_g/N_q = -8$ (curve~3),
$G_1$ passes through zero at $z \sim 10^{-3}$. When $N_g/N_q <
-10$, $G_1$ is positive at any experimentally reachable $z$
(curve~4)~. Therefore, the
 experimental measurement of the turning point
would allow to draw conclusions on the interplay between the
initial quark and gluon densities.
\section{Remark on the higher twists contributions}
In the region \textbf{B} one can expand terms $\sim (Q^2 +
\mu^2)^k$ in Eq.~(\ref{gnsbc}) into series  in $(\mu^2/Q^2)^n$ and
represent $g_1^{NS}(x+z,Q^2,\mu^2)$ as follows:

\begin{equation}\label{gnshtw}
g_1^{NS}(x+z,Q^2,\mu^2)= g_1^{NS}(x,Q^2/\mu^2) + \sum_{k=1}
(\mu^2/Q^2)^k T_k~
\end{equation}
where $g_1^{NS}(x,Q^2/\mu^2)$ is given by Eq.~(\ref{gnsint});
 for explicit expressions for the factors $T_k$ see
Ref.~\cite{egthtw}. The power terms in the  rhs of
Eq.~(\ref{gnshtw}) look like the power $\sim 1/(Q^2)^k$
-corrections and therefore the lhs of Eq.~(\ref{gnshtw})  can be
interpreted as the total resummation of such corrections. These
corrections are of the perturbative origin and have nothing in
common with higher twists contributions ($\equiv HTW$). The latter
appear in the conventional analysis of experimental date on the
Polarized DIS as a discrepancy between the data and the
theoretical predictions, with $g_1^{NS}(x,Q^2/\mu^2)$ being given
by the Standard Approach:
\begin{equation}\label{dglaphtw}
g_1^{NS~exp} = g_1^{NS SA} + HTW~.
\end{equation}
Confronting Eq.~(\ref{dglaphtw}) to Eq.~(\ref{gnshtw}) leads to an
obvious conclusion: For estimating genuine higher twists
contributions to $g_1^{NS}$, one should account, in the first
place, for the perturbative power corrections predicted by
Eq.~(\ref{gnshtw}); otherwise the estimates cannot be reliable. It
is worth mentioning that we can easily explain the empirical
observation made in the conventional analysis of experimental
data: The power corrections exist for $Q^2 > 1$ GeV$^2$ and
disappear when $Q^2 \to 1$ GeV$^2$. Indeed, in Eq.~(\ref{gnshtw})
$\mu = 1$ GeV , so the expansion in the rhs of Eq.~(\ref{gnshtw})
make sense for $Q^2 > 1$ GeV$^2$ only; at lesser $Q^2$ it should
be replaced by the expansion of Eq.~(\ref{gnsbc}) in
$(Q^2/\mu^2)^n$.
\section{Conclusion}

Resummation of the leading logarithms of $x$ is the
straightforward and most natural way to describe $g_1$ at small
$x$. Contrary to DGLAP, our approach is not sensitive to the value
of $Q^2$ and allows one to describe $g_1$ at small $x$ and
arbitrary $Q^2$ in terms of the same expressions at large and
small $Q^2$. We have used it for studying the $g_1$ singlet at
small $Q^2$ because this kinematic is presently investigated by
the COMPASS collaboration. It turns out that $g_1$ in this region
depends on $z = \mu^2/2pq$ only and practically does not depend on
$x$, even at  $x \ll 1$. Numerical calculations show that the sign
of $g_1$ is positive at $z$ close to 1 and can remain positive or
become negative at smaller $z$, depending on the ratio between
$\delta g$ and $\delta q$. It is plotted in Fig.~1 for different
values of $\delta g/\delta q$. Fig.~1 demonstrates that the
position of the sign change point is sensitive to the ratio
$\delta g/\delta q$, so the experimental measurement of this point
would enable to estimate the impact of $\delta g$.

The alternative to the resummation is extrapolating DGLAP from its
natural region of applicability (large $x$ and large $Q^2$) into
the region of small $x$ and large $Q^2$. As the DGLAP equations
cannot account  for the LL resummation, SA mimics the resummation
through the special choice of the fits for the initial parton
densities: the singular factors in the fits cause the steep rise
of $g_1$ at small $x$ and provide the Regge asymptotics for $g_1$
(however with the incorrect phenomenological intercepts) when $x
\to 0$. They should be dropped when the total resummation of LL of
$x$ is taken into account. The remaining, regular $x$-terms of the
DGLAP fits (the function $\varphi$ in Eq.~(\ref{fits})) can
obviously be replaced by much simpler expressions, so the number
of phenomenological parameters in the fits can be reduced from 5
to 2 or even 1. To conclude, let us notice that extrapolating
DGLAP into the small-$x$ region, though provides a satisfactory
agreement with experimental data, leads to various wrong
statements. We enlisted the most of them in a recent
Ref.~\cite{egtep}. Below we mention two more such wrong
 statements:

\textbf{Statement 1:} \emph{The $Q^2$-power corrections stem from
higher twists $g_1$ and can be measured as the discrepancy between
the DGLAP predictions and the data.} This statement is wrong as
shown in the previous Sect.

\textbf{Statement 2:} \emph{The impact of the LL resummation on
the small-$x$ behavior of $g_1$ is small.} This statement appears
when the resummation has been included into the DGLAP expressions
where the fits contain singular factors. Such inclusion  is
inconsistent and  means actually a double counting  of the  LL
contributions: once through the fits and secondly in the explicit
way. It also affects  the small-$x$ asymptotics of $g_1$, leading
to the incorrect values of the intercepts of $g_1$ (see
Ref.~\cite{egtfit} for more detail).

\section{Acknowledgement}
B.I.~Ermolaev is grateful to the Organizing Committee of the
workshop DIS 2007 for financial support of his participation in
the workshop.

\end{document}